\documentclass[aps,llncs,preprint]{revtex4-1}
\usepackage{graphicx}
\usepackage{amsmath}
\usepackage{amssymb}
\usepackage{float}
\usepackage{enumitem}
\usepackage{subcaption}
\usepackage{color}
\DeclareUnicodeCharacter{2212}{-}
\begin{document}

\title{Andreev reflection spectroscopy on SnAs  single crystals}

\author{Sandeep Howlader$^1$}
\author{Nikhlesh Mehta$^1$}
\author{M. M. Sharma$^2$}
\author{V.P.S. Awana$^3$}

\author{Goutam Sheet$^1$}

\email{goutam@iisermohali.ac.in}
\affiliation{$^1$Department of Physical Sciences, 
Indian Institute of Science Education and Research Mohali, 81, Knowledge city, S.A.S. Nagar, Manauli 140306, Punjab, India}
\affiliation{$^2$ Academy of Scientific \& Innovative Research (AcSIR), Ghaziabad-201002}
\affiliation{$^3$ CSIR- National Physical Laboratory, New Delhi-110012}
\begin{abstract}

Binary compounds of SnT family(T= As, Sb, Te, S, Se, P) exhibit novel properties like superconductivity, topologically protected states etc. Some of these compounds crystallize in NaCl type structure or in layered structure, both of which are considered potentially important for high-temperature/unconventional superconductivity.  It was previously shown that K-doped BaBiO$_3$ with NaCl type structure exhibits superconducting state below $\sim$ 30 K. SnAs  has  crystallographic configuration that is exactly similar to K-doped BaBiO$_3$ and this warrants for the investigation of superconductivity in this material. Previously it was reported that SnAs exhibits weakly coupled type I superconductivity with an energy gap of 0.7 meV. Recently, the electronic bandstructure calculations has hinted to the existence of possible topologically protected states in SnAs. This has motivated us to investigate the superconducting nature of SnAs using point contact Andreev reflection spectroscopy. Our investigation revealed that superconductivity in SnAs can be well explained within the BCS framework in the weak coupling limit.
\end{abstract}
\maketitle

Recently, investigation of the electronic properties of binary compounds of SnT family (T= As, Sb, Te, S, Se, P) has garnered a lot of attention\cite{Zhao,Sun,Kamitani,Hsieh}. These materials exhibit novel properties and are considered as sister compounds of topological materials (e.g. topological semimetals and topologcial crystalline insulators), while some were also found to be superconductors\cite{Bezotosnyi}. The SnT family of compounds are also candidate materials for topological superconductivity. In the context of superconductivity, In doped SnTe was found to host an unusual pairing mechanism\cite{Erickson, Balakrishnan, Novak}.  Among the members of this family of materials, being host to anisotropic properties, the materials with NaCl-type structure are most widely studied\cite{Hulliger, Tutuncu, Wang, Tanaka1, Tanaka2}. Even though the NaCl-type structural compounds are not layered, due to highly symmetric nature of NaCl-type SnT, it is possible that due to the presence of a single crystallographically independent site for Sn, the typical mixed valence state of Sn (Sn$^{4+}$(5s$^0$) and Sn$^{2+}$(5s$^2$)) becomes naturally forbidden. This is quite similar to the case of BaBiO$_3$ where upon doping with Potassium a high temperature superconducting phase was obtained\cite{Cava, Mattheiss}. Therefore it becomes highly probable that NaCl-type SnT materials may exhibit high T$_c$ superconductivity. 

Tin arsenide(SnAs) is a material that belongs to the SnT family and crystallizes in the NaCl-type structure. SnAs was first demonstrated to exhibit superconductivity by Geller and Hull in 1964\cite{Geller}. In 2014, Wang et. al performed transport and heat capacity measurements in SnAs and  reported that SnAs exhibits weakly coupled type-I superconductivity ($\lambda = 0.62$) with a low transition temperature of 3.58 K\cite{Wang1}. Additionally,  instead of being in a mixed valence state, Sn was found to exist in a single valence state. Phonon dispersion curves and phonon DOS were studied thoeretically which revealed that the As p-states hybridize strongly with Sn p-states at the Fermi level\cite{Tutuncu1}. Moreover, DFT calculations  revealed that spin-orbit coupling does not have any significant effect on the bandstructure. However, ARPES measurements revealed significant band splitting\cite{Bezotosnyi} . More recent investigation of electronic band structure and Density of States(DOS) within the Density Functional Theory (DFT) posited the possibility of  non-trivial topology in SnAs bands. Furthermore, the band dispersion obtained for SnAs were also found to closely resemble to that of SnTe, a topological crystalline insulator\cite{Sharma}. These calculations also revealed  band inversion, gapped states near the Fermi level and possible existence of topological surface states. This makes SnAs a good system for studying possible topological superconductivity. In this paper, we investigated the possibility of unconventional superconductivity in SnAs by performing Andreev reflection spectroscopy at low temperatures and under the presence of high magnetic fields.  \\
\begin{figure}[h!]
\centering
\includegraphics[scale=0.5]{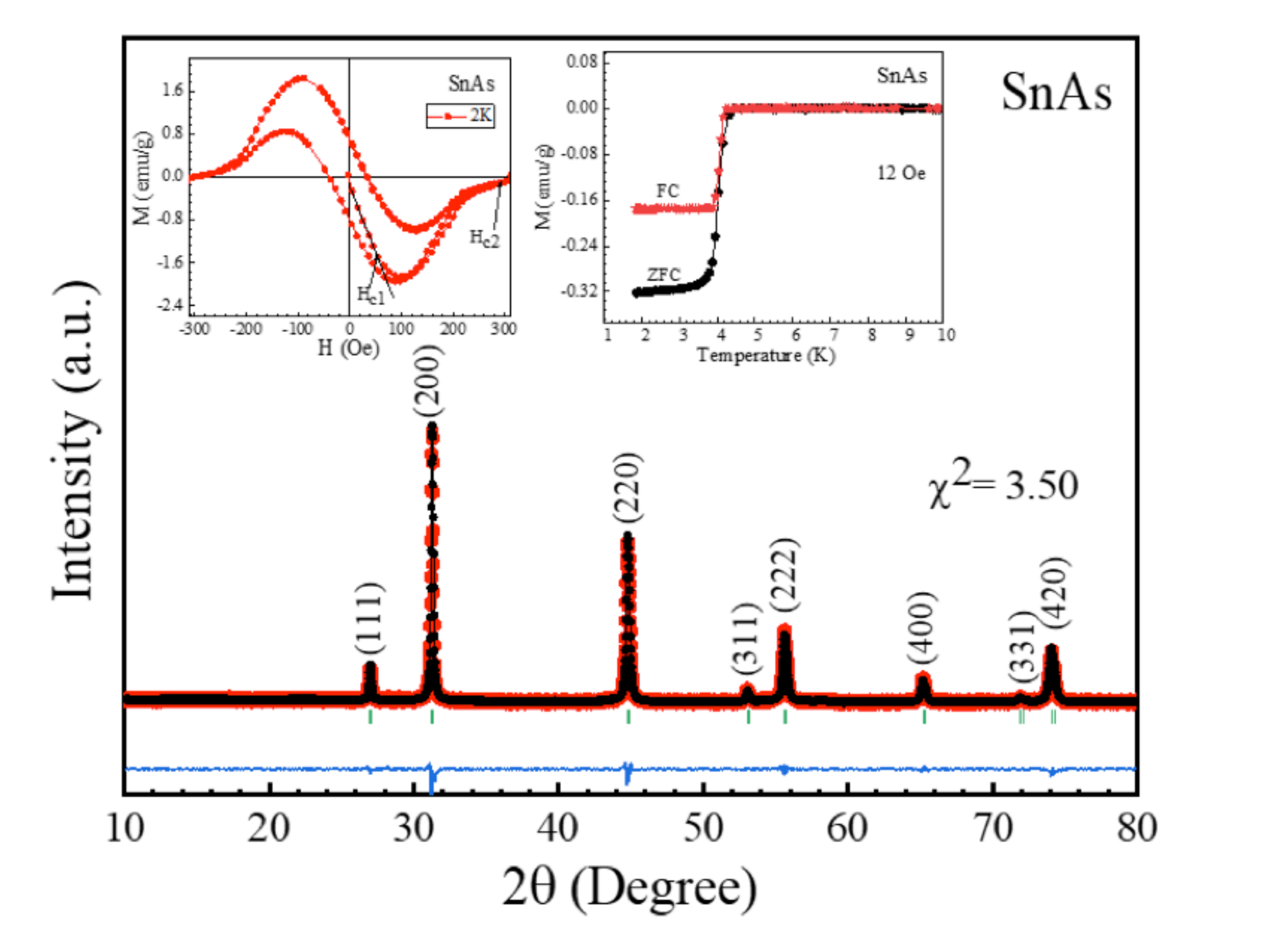}
\caption{ Rietveld refined XRD pattern of synthesized SnAs crystal, right inset is showing FC and ZFC measurements of SnAs and left inset is showing M-H plot at 2K of synthesized SnAs crystal\cite{Sharma}. }
%\label{crkt}
\end{figure}

The SnAs crystals used in this study were grown by following a two-step method as described in our previous report\cite{Sharma}. Phase purity of the synthesized SnAs crystal was determined through XRD pattern. Rietveld refined XRD pattern is shown in Fig.1. Quality of refinement is determined by $\chi^2$ (goodness of fit parameter) value which is found to be 3.50. The synthesized SnAs crystal has NaCl type cubic structure with Fm-3m space group symmetry. Sn and As occupy (0,0,0) and (0.5,0.5,0.5) atomic position in SnAs crystal lattice. The lattice parameters obtained by reitveild refinement are shown in Table (1). No impurity peak is visible in XRD pattern, showing that the synthesized crystal is phase pure. Magnetization measurements under field-cooled (FC) and zero field-cooled (ZFC) protocols were performed in presence of 12Oe magnetic field. Both of these measurements confirmed the presence of bulk superconductivity in the form of diamagnetic transition at around 4K as shown in right inset of Fig. 1. This diamagnetic signal seems to saturate near 3.8 K, showing a transition width of 0.2K. Magnetization vs applied Field (M-H) plot at 2K is shown in left inset of Fig.1. This wide open M-H loop shows the observed superconductivity in synthesized sample is type-II superconductivity. All previous reports shows SnAs to be a type-I superconductor with a critical field of 150Oe \cite{Bezotosnyi, Wang}. In this result M-H loop starts to deviate from linearity from around 60 Oe and it closes at around 290 Oe. This result shows the superconductivity in synthesized SnAs crystal to be type-II superconductivity. This result is discussed in detail in our previous report \cite{Sharma}. \\

\begin{table}[ht]

\centering
\begin{tabular}{c c c c c c c }
\hline\hline

\hline
Sample & a(\AA) & b(\AA) & c(\AA) & $\alpha$ & $\beta$ & $\gamma$ \\
SnAs & 5.721(9) & 5.721(9) & 5.721(9) & 90 & 90 & 90\\

\hline

\end{tabular}
\caption{Rietveld refined lattice parameters of SnAs}
\label{table:nonlin}
\end{table}

\begin{figure}[h!]
%\centering
\includegraphics[scale=0.75]{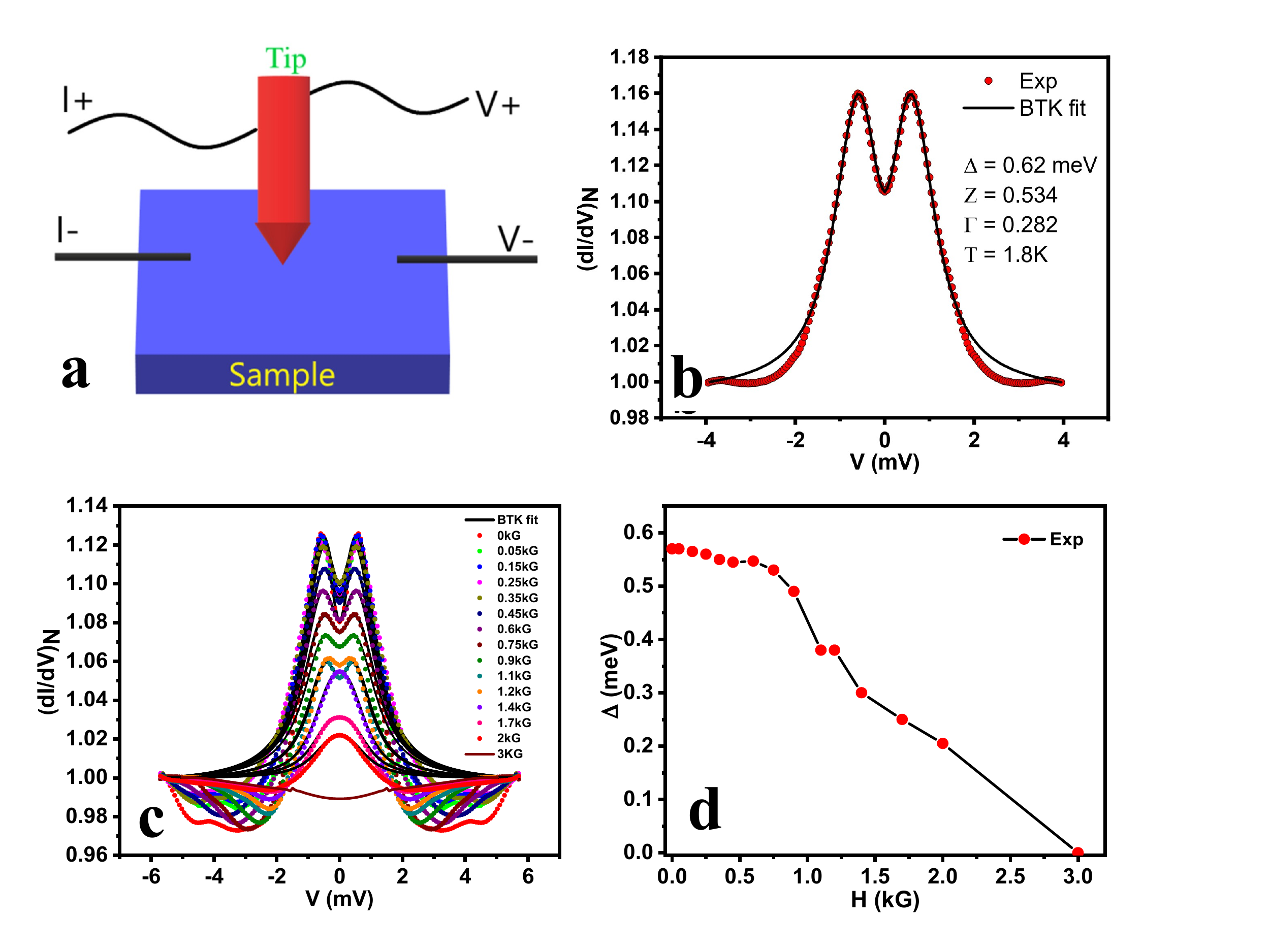}
\caption{a) Needle-anvil technique of point contact formation (b) a differential conductance spectrum obtained in the ballistic regime (c) Field dependence of dI/dV vs. V spectrum along with BTK-simulated curves (d) $\Delta$ vs. H }
%\label{crkt}
\end{figure}
The point contact Andreev reflection spectroscopy measurements were performed in a liquid helium cryostat which is equipped with a variable temperature insert(VTI) operational between 1.4 K and 300 K. The cryostat is also  equipped with a superconducting vector magnet(6T-1T-1T). A home-built probe which utilizes the needle anvil method, was used to form point contacts at low temperatures.  Here, a mesoscopic point contact junction was formed by engaging a silver(Ag) tip on the surface of SnAs crystal. Contacts where the diameter were small yielded two peaks in dI/dV vs V spectra, symmetric about the zero bias. These peaks are signature of Andreev reflection phenomenon\cite{Duif, Sheet, Jansen}. 
Andreev reflection process is observed in ballistic point contacts between a normal metal and a superconductor that leads to a special type of non-linearity in the I-V spectrum. For such point contacts, the non linearity is directly probed in a dI/dV vs. V spectrum. Andreev reflection spectra are analyzed via the Blonder, Tinkham, and Klapwijk(BTK) model\cite{BTK, Strijkers}. Here, the  normal metal/superconductor (N/S) interface is modeled as a $\delta$-function potential barrier and its strength is characterized by $Z$, a dimensionless parameter. For elemental superconductors, the quasi-particle life-time (represented by $\Gamma$) is very small. Andreev reflection spectra obtained for mesoscopic interfaces for such superconductors can be fitted by using just two fitting parameters, the barrier potential and the superconducting energy gap ($\Delta$). For non-zero barrier potential ($Z > 0$), the differential conductance spectrum obtained for a ballistic contact contains Andreev peaks at bias V = $\pm \Delta$/e. 

\begin{figure}[h!]
%\centering
\includegraphics[scale=0.65]{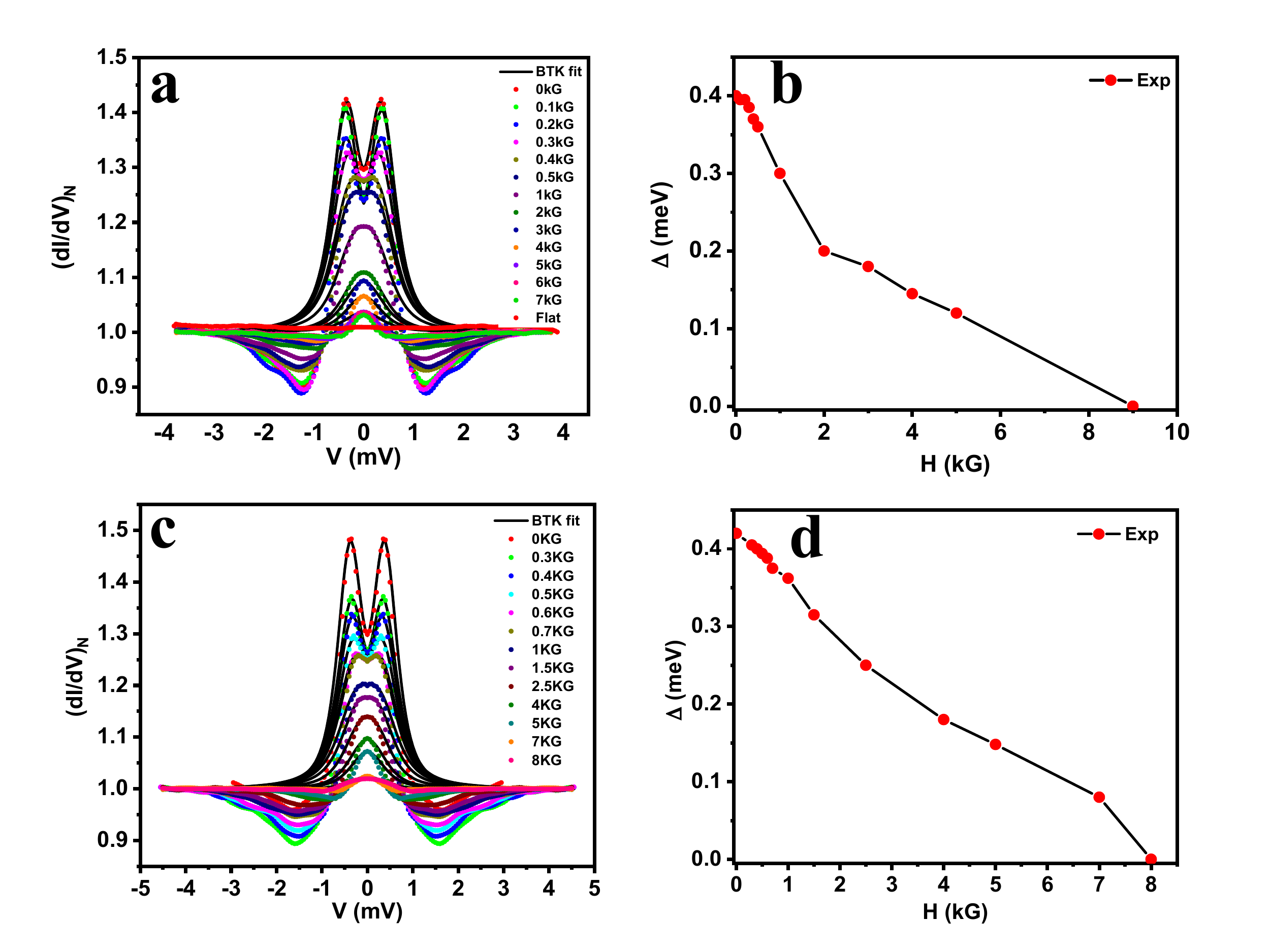}
\caption{a), (c) Magnetic field dependence of point contact spectra along with BTK-simulated curves (b), (d) corresponding $\Delta$ vs. H  plots. }
%\label{crkt}
\end{figure}

In Fig. 2a we show a typical point-contact formed using the needle-anvil method. The corresponding normalized dI/dV vs. V spectrum obtained for the contact is shown in Fig 2b.  The spectrum was fitted under the BTK formalism and corresponding fitting parameters used are shown in the figure.  It is clearly seen that the spectrum is well-fitted in the range $\pm$ 2 mV. Additionally, the spectrum also contains two dips in dI/dV at a higher bias arising due to critical current dominated non-linearities in I-V.  Differential conductance spectra obtained for a purely ballistic contact can be fitted in the entire range under the BTK formalism. However, additional features due to non-linearities in I-V that arise due to factors related to scattering mechanisms occuring within the contact cannot be fitted within the BTK theory and leads to under-estimation of superconducting energy gap ($\Delta$). By fitting the Andreev reflection spectra we estimate $\Delta$ to be $\sim$ 0.62 meV. This is in accordance with the results obtained by Bezotosnyi et al\cite{Bezotosnyi}.\\

In order to gain further understanding of the superconducting phase, detailed field and temperature dependent measurements were carried out. We applied magnetic field along the axis of the point contact ($B || c$).  The dI/dV spectra evolved smoothly with applied magnetic field (shown in Fig. 2c). The critical field for a superconducting point contact is the applied field strength for which the peak-dip structure completely vanishes. For our case this happened at 3 kG.  All of the spectra  were fitted according to the BTK formalism and information regarding change of energy gap with field was extracted. The spectra were normalized to the conductance at the highest bias value. Additionally, with increasing magnetic field the  critical current dip structure was observed to shift inwards. We present the change in $\Delta$ with applied magnetic field in Fig. 2d where, the energy gap is seen to decrease monotonically with applied field strength.\\

We also studied field dependent measurements on some of the other point contacts, where critical current dominated effects were prominent. These measurements are shown in Fig. 3. Both spectra were seen to evolve monotonically with applied magnetic field. These spectra were also fitted according to the BTK theory and were normalized to the conductance at highest bias value. The estimated value of $\Delta$ for these contacts is lower than the estimated value for spectra presented in Fig. 2b. This reduction is a result of analysing a spectrum where critical current dominated features are present. As we have mentioned before, such features lead to underestimation of energy gap. The overall spectroscopic features for these point contacts vanished at a critical field of 8 kG. \\

\begin{figure}[h!]
%\centering
\includegraphics[scale=0.65]{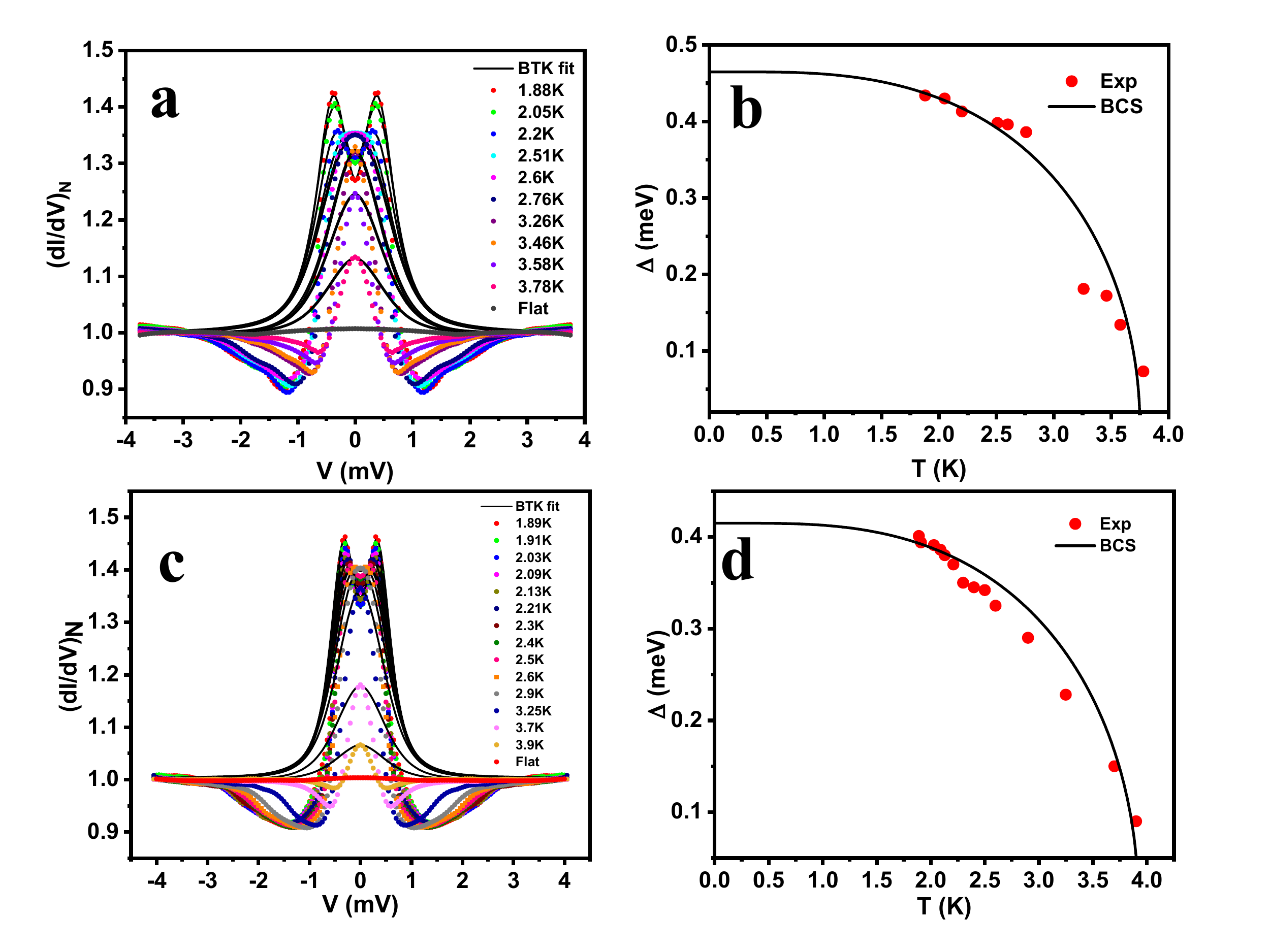}
\caption{a), c) Temperature dependence of SnAs/Ag point contact spectra along with BTK-simulated curves b),d) corresponding $\Delta$ vs. T with BCS simulated curve.  }
%\label{crkt}
\end{figure}

We present the temperature dependent measurements in Fig. 4. Here, we show temperature dependent data for point contacts where both, spectroscopic signatures associated with Andreev reflection and critical current of point contact, are present.  The data is normalized to the conductance at the highest bias value. As expected for superconductors, the spectra evolved smoothly with increasing temperature and the overall spectroscopic features disappeared above 4.2 K.  Again, as in the case of field dependence, the spectroscopic features appearing due to critical current (dips in dI/dV vs. V) shift towards lower bias value. This indicates temperature dependent suppression of critical current for the point contacts. \\

Superconducting energy gap and its variation with temperature was obtained by fitting the data under the BTK framework.  We have shown the temperature variation of $\Delta$ in Fig. 4b and 4d. The dots represent the estimated value of $\Delta$ from BTK fitting of experimental data and the solid line represent the simulated  curve following the BCS prediction\cite{Gasperovic, Bardeen}. Clearly the obtained variation of $\Delta$ with temperature follows the BCS prediction. Furthermore, the value of $\Delta$ as determined by BTK simulation provides the ratio $\Delta(0)/K_BT_C = 1.78$ indicating that superconductivity in SnAs can be explained by BCS theory in the weak coupling limit. Additionally, we did not detect any features related to unconventional nature of superconducting order parameter. \\

In conclusion, we have performed spectroscopic investigation of superconductivity in NaCl crystallized SnAs single crystals. We have estimated the superconducting energy gap to 0.62 meV. On the basis of detailed temperature and field dependence of various point contacts on SnAs, and corresponding BTK simulated curves, we conclude that superconductivity in SnAs is well explained within the BCS framework. We have determined that the critical temperature is 4.2 K and critical field is 8 kG. Based on our spectroscopic investigation we conclude that SnAs is a conventional superconductor. 

G. S. would like to acknowledge financial support from Swarnajayanti fellowship awarded by the Department of Science and Technology (DST), Govt. of India (grant No. DST/SJF/PSA-01/2015-16).

\end{document}